\documentclass[lettersize,journal]{IEEEtran}
\usepackage{amsmath,amsfonts}
\usepackage{algorithmic}
\usepackage[ruled,linesnumbered]{algorithm2e}
\usepackage{amssymb}
\usepackage{array}
\usepackage{color}
\usepackage{multirow}
\usepackage[caption=false,font=normalsize,labelfont=sf,textfont=sf]{subfig}
\usepackage{textcomp}
\usepackage{stfloats}
\usepackage{url} 
\usepackage{verbatim}
\usepackage{graphicx}
\usepackage{threeparttable}
\usepackage{cite}
\usepackage[colorlinks]{hyperref}
\usepackage{balance}
\begin{document}

\title{Probabilistic ODMA Receiver with Low-Complexity Algorithm for MIMO Unsourced Random Access}
\author{Zhentian Zhang,~\IEEEmembership{Student Member,~IEEE,} Jian Dang,~\IEEEmembership{Senior Member,~IEEE,}
	Zaichen Zhang,~\IEEEmembership{Senior Member,~IEEE}
	\thanks{
		Z. Zhang, J. Dang, and Z. Zhang are with the National Mobile Communications Research Laboratories of Southeast University, Nanjing, China. Jian Dang and Zaichen Zhang are the corresponding authors.
	}
}
%\author{IEEE Publication Technology,~\IEEEmembership{Staff,~IEEE,}
%        % <-this % stops a space
%\thanks{This paper was produced by the IEEE Publication Technology Group. They are in Piscataway, NJ.}% <-this % stops a space
%\thanks{Manuscript received April 19, 2021; revised August 16, 2021.}}



\maketitle

\begin{abstract}
	In this work, we present the design for both pilot-uncoupled and pilot-free on-off multiple access (ODMA) receivers in unsourced random access (URA) for multiple-input multiple-output (MIMO) systems. Unlike pilot-coupled ODMA, where on-off patterns are linked to pilot selection, pilot-uncoupled and pilot-free ODMA reduce transmission redundancy but face challenges in processing complexity and capacity performance. The joint pattern and data detector (JPDD) design is critical for these schemes, but the current JPDD algorithm has high complexity with quadratic computational costs. To address this, we propose a low-complexity detector based on approximate message passing (AMP), which offers linear complexity, providing reduced cost and improved performance in the under-determined linear regression case. Decoding is initialized via pilot-free matrix factorization through alternating minimization, resolving phase and scalar ambiguities. Compared to existing pilot-free schemes, the proposed method achieves a 13 dB improvement and favorable trade-offs in complexity and capacity performance when compared to benchmarks.
\end{abstract}


\section{Introduction}
\subsubsection{Background And Related Works}
Future demands for massive connectivity necessitate advanced multiple access techniques that provide lower latency and higher system capacity\cite{WYP_Survey}. Currently, scenarios such as massive machine-type communication (mMTC) are extensively analyzed in the context of finite block length and short packet transmission\cite{Massive_Access}. Conventional multiple access techniques, such as ALOHA, often prove inefficient and struggle to handle the increased device density. Fortunately, unsourced random access (URA) offers promising insights for future system designs\cite{Polyansky1}. URA redefines multiple access as a theoretical coding problem, yielding elegant analytical capacity results. Since its introduction, extensive research has focused on achieving the random coding bound using a specific coding scheme\cite{URA_China_Comm}.

On-off division multiple access (ODMA)-based URA demonstrates great flexibility and scalability, making it a strong candidate for future applications\cite{ODMA1,ODMA2,ODMA3,ODMA4}. ODMA encodes binary messages into two coupled parts. One portion of the data (typically a dozen of bits) selects the on-off pattern from a common codebook, while the other portion is transmitted using the corresponding transmission pattern. Both the pattern and the permuted data are typically detected and restored jointly at the receiver side.
\subsubsection{Challenges}
Regarding multiple input and multiple output (MIMO) systems, the MIMO-ODMA receiver has been explored from various perspectives with different methods of pattern bit projection\cite{ODMA5,IOTJ_ODMA,DL_ODMA,TWC_ODMA}. Currently, there are three existing categories: pilot-coupled, pilot-uncoupled, and pilot-free. In the pilot-coupled approach\cite{ODMA5}, the on-off pattern is linked to pilot selection, meaning the transmission pattern is restored simultaneously after detecting activity in the pilot codewords. In the pilot-uncoupled approach\cite{IOTJ_ODMA}, the pilot codewords and on-off patterns are selected by different bit sequences, and the restoration of transmission patterns is performed jointly when the receiver decodes the transmitted data. In contrast, the pilot-free approach\cite{DL_ODMA,TWC_ODMA} directly performs pattern and data detection via matrix factorization using a dictionary designed by channel model statistical features.

While the pilot-uncoupled and pilot-free schemes can save more redundancy (i.e., fewer bits are input into the channel code, resulting in fewer parity check bits and less modulated constellation transmission), they impose higher demands on signal processing. A more stringent issue is that there still exists a significant capacity gap, with the pilot-free schemes performing worse compared to the pilot-based approaches. Additionally, the complexity of the current multi-user detector for the pilot-uncoupled scheme remains quadratic, which limits the feasibility of iterative decoding designs due to its intolerable complexity.

\subsubsection{Contributions}
In this work, we propose an ODMA-based URA with a probabilistic\footnote{We use the term 'probabilistic' to refer to the blind detection involved in matrix decomposition and the calculation of a posteriori probabilities at the receiver side.} receiver enabled by alternate matrix factorization. While it retains the framework of current pilot-free schemes (e.g., decoding with phase and scalar ambiguities), it allows for more accurate channel information to be obtained, rather than relying solely on statistical channel information. Furthermore, compared to the pilot-uncoupled scheme, the complexity of joint pattern and data detection (JPDD) is reduced from quadratic to linear order

The specific contributions are as follows:
\begin{itemize}
	\item[-] An ODMA-based URA scheme for MIMO systems with slotted transmission is designed. Similar to the pilot-free receiver structure, decoding is initialized through matrix factorization. Instead of methods such as singular value decomposition, the low-rank feature of the signal component is exploited using alternate minimization decomposition.
	\item[-] A novel JPDD multi-user detector, inspired by approximate message passing (AMP), is proposed for pilot-uncoupled ODMA in a posteriori probability (APP) calculation. It requires only linear complexity and incurs only marginal performance loss, making pilot-uncoupled ODMA a more practical scheme. Numerical results also show that the performance of JPDD can be improved in the under-determined linear regression case, as no matrix inverse calculations are required.
	\item[-] The capacity performance of the pilot-free receiver is improved by up to 13 dB, with a minimum improvement of 8 dB. Although there is still a gap (ranging from 0.5 dB to 4 dB) compared to the state-of-the-art methods, a good trade-off between capacity and complexity can be observed.
\end{itemize}
\subsubsection{Content Structures}The rest of the paper is organized as follows: Section~\ref{sec.2} describes the system model for ODMA in MIMO URA. Section~\ref{sec.3} details the encoder design in the proposed scheme. Section~\ref{sec.4} elaborates on the major components of the proposed probabilistic receiver. Section~\ref{sec.5} presents numerical results with fair comparisons to state-of-the-art methods. Finally, conclusions and future research directions are discussed in Section~\ref{sec.6}.
\section{System Descriptions}\label{sec.2}
In this work, there are $K_a$ single antenna devices to be served by an $M$-antenna receiver. Each device transmits information of $B$ bits within the channel coherence time over $L$ channel uses split into $J$ chunks with $T$ length each, i.e., $T=L/J$. Let $\boldsymbol{u}_k \in \{0,1\}^{B\times 1}, k \in \{1,2,\ldots,K_a\}$ denote the binary message vector of the $k$-th device and set $\mathcal{L}$ denote all binary messages. After encoding procedures, the frame to be transmitted is denoted by $\boldsymbol{x}_k \in \mathbb{C}^{T\times 1}$. The multiple access channel vector is denoted by $\boldsymbol{h}_k \in \mathbb{C}^{M \times 1}$. This work considers the URA channel model with  $\boldsymbol{h}_k \sim \mathcal{CN}(0,\boldsymbol{I})$. Thereupon, the received signal $\boldsymbol{Y} \in \mathbb{C}^{M\times T}$ at single chunk is given by:
\begin{equation}
	\label{eq:1}
	\begin{aligned}
		\boldsymbol{Y} =\sum_{k=1}^{\hat{K}_a}\boldsymbol{h}_k\boldsymbol{x}_k^{\mathrm{T}}+\boldsymbol{N} =\boldsymbol{H}\boldsymbol{X}+\boldsymbol{N},
	\end{aligned}
\end{equation}
where the number of active devices at single chunk is denoted by $\hat{K}_a \ll K_a$, $\boldsymbol{N}$ is the additive white Gaussian noise with zero mean and variance $\sigma^2$, i.e., $\boldsymbol{N} \sim \mathcal{CN}(0,\sigma^2\boldsymbol{I})$, $\boldsymbol{H} \in \mathbb{C}^{M \times \hat{K}_a}$ is the channel matrix and $\boldsymbol{X} \in \mathbb{C}^{\hat{K}_a \times T}$ contains the frames of all $\hat{K}_a$ devices at its rows. Intuitively, with appropriate frame length setup, the signal component in \eqref{eq:1} features low matrix rank by $rank(\boldsymbol{HX})\le \hat{K}_a \ll M$ and $rank(\boldsymbol{HX})\le \hat{K}_a \ll T$. By exploiting such feature, this work proposes a probabilistic receiver design with low-rank matrix factorization tailored for ODMA of MIMO URA. The system energy constraint is entitled as energy-per-bit $E_b/N_0=\frac{P}{\sigma^2B}$ where $P$ is the total transmission power. The goal at the receiver is to produce a list $\tilde{\mathcal{L}}$ with $K_a$ messages from all $J$ chunks. The system performance is evaluated by per-user probability of error (PUPE) denoted by $P_e$ which is the summation of missed detection and false alarm, consisting of:
\begin{equation}
	\label{eq:2}
	P_{\mathrm{md}}=\frac{\mathbb{E}[n_{\mathrm{md}}]}{K_a},
P_{\mathrm{fa}} =\mathbb{E}\left[\frac{n_{\mathrm{fa}}}{|\mathcal{\tilde{L}}|}\right],
\end{equation}
where the size of output list is $|\mathcal{\tilde{L}}|=K_a-n_{\mathrm{md}}+n_{\mathrm{fa}}$. 
\begin{figure}[htp]
	\centering
	\includegraphics[width=2.8in]{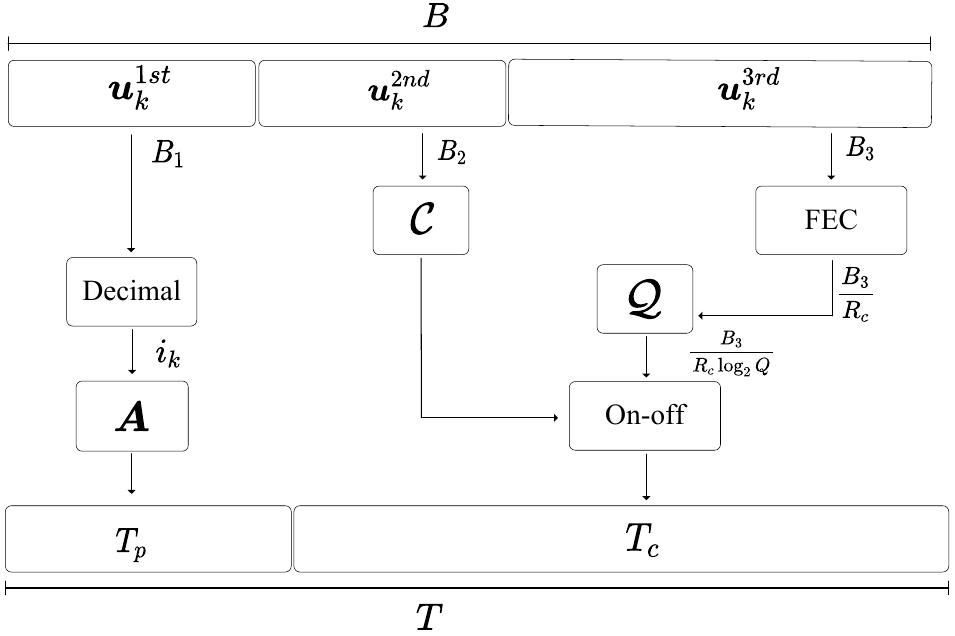}
	\caption{Illustration of transmission in single chunk and the proposed encoder procedures where $\mathcal{Q}$ denotes modulation with order $Q$, $\boldsymbol{A}$ and $\mathcal{C}$ denote pattern codebooks. The $B_2$ bits are used to select patterns only and not transmitted at all and the chunk has length equal to $T=L/J$}.
	\label{encoder}
\end{figure}
\section{Encoder Design}\label{sec.3}
This subsection elaborates the encoding procedures in this work. The goal is to improve transmission efficiency by utilizing on-off patterns to carry information. Overall, the binary messages undergo segmentation, forward error correction (FEC) encoding with code rate $R_c < 1$, modulation with order $Q$ and permutation. Eventually, a sparse frame $\boldsymbol{x}_k$ is generated consisting of two consecutive sub-frames with length $T_p$ and $T_c$ respectively, where $T=T_p+T_c$. The encoding procedures are illustrated in Fig. \ref{encoder}.

Initially, the message vector $\boldsymbol{u}_k$ of the $k$-th active device is divided into several segments: the decimal expression of the first $B_{\mathrm{chunk}}$ bits are used to select the chunk for transmission, $\boldsymbol{u}_k^{1st}=\boldsymbol{u}_k\left(B_{\mathrm{chunk}}:B_1+B_{\mathrm{chunk}},1\right),\boldsymbol{u}_k^{2nd}=\boldsymbol{u}_k\left(B_1+B_{\mathrm{chunk}}+1:B_1+B_{\mathrm{chunk}}+B_2,1\right),\boldsymbol{u}_k^{3rd}=\boldsymbol{u}_k\left(B_1+B_{\mathrm{chunk}}+B_2+1:B,1\right)$ with $B_1$, $B_2$ and $B_3$ bits respectively, where $B_{\mathrm{chunk}}+B_1+B_2+B_3=B$. Meanwhile, there are two codebooks with different sizes, namely $\boldsymbol{A}= \left[\boldsymbol{a}_{1},\boldsymbol{a}_{2},\ldots,\boldsymbol{a}_{2^{B_1}}\right] \in \mathbb{C}^{T_p \times 2^{B_1}}$ and $\mathcal{C} \in \{0,1\}^{T_c \times 2^{B_2}}$. For codebook $\boldsymbol{A}$, it's commonly generated from sub-sampled DFT matrix or Gaussian signatures with power constraint $\alpha P$ where $\alpha$ denote the power allocation ratio. For codebook $\mathcal{C}$, the codewords have only $\frac{B_3}{R_c \log_2Q}$ elements equaling to one and others all zero. For modulation, set $\mathcal{Q}$ contains all constellation symbols.

The first segment $\boldsymbol{u}_k^{1st}$ is encoded into the first $T_p$ time of slots. Device selects a codeword from $\boldsymbol{A}$ by the decimal integer $i_{k}\in \{1,2,\ldots,2^{B_1}\}$ (radix-2 plus one) of the binary segment $\boldsymbol{u}_k^{1st}$, i.e., $i_k = \left[\boldsymbol{u}_k^{1st}\right]_2+1$. Subsequently, segment $\boldsymbol{u}_k^{1st}$ is non-linearly modulated into codeword in $\boldsymbol{A}$. Moreover, the pattern of the later $T_c$ time of slots is selected by the integer value of segment $\boldsymbol{u}_k^{2nd}$ from codebook $\mathcal{C}$. Segment $\boldsymbol{u}_k^{3rd}$ is firstly FEC encoded and then modulated into constellation vector $\boldsymbol{s}^{2nd}_k \in \mathbb{C}^{\frac{B_3}{R_c\log_2Q} \times 1}$. FEC codes can be arbitrary with feasible modifications such as low density parity check (LDPC) codes, Polar code and etc. Then, vector $\boldsymbol{s}^{2nd}_k$ is permuted by the selected on-off pattern from $\mathcal{C}$ into frames $\boldsymbol{x}_k(T_p+1:T,1)$. Thus, the frame to be transmitted is given by $\boldsymbol{x}_k= \left[\boldsymbol{a}_{i_k}^{\mathrm{T}}, \boldsymbol{x}_k(T_p+1:T,1)^{\mathrm{T}}\right]^{\mathrm{T}}$.
%\begin{equation}
%	\label{eq:3}
%	\boldsymbol{x}_k= \left[\boldsymbol{a}_{i_k}^{\mathrm{T}}, \boldsymbol{x}_k(T_p+1:T,1)^{\mathrm{T}}\right]^{\mathrm{T}}.
%\end{equation}
Notably, the information of segment $\boldsymbol{u}_k^{2nd}$ is embedded into the pattern of the frames $\boldsymbol{x}_k(T_p+1:T,1)$, i.e., these $B_2$ bits are not transmitted but can be restored at the receiver.
\section{Decoder Design}\label{sec.4}
This section elaborates the proposed decoder exploiting features of the low-rank matrix and the sparsity in active on-off patterns. Five major components work in tandem, namely, activity detection by power information, low-rank matrix factorization via alternate minimization, ambiguity conpensation and joint pattern and data detection, successive interference cancellation (SIC).
\subsubsection{Estimation By Power Information}
Normally, the number of activity $\hat{K}_a$ at single chunk is unknown due to randomness. Therefore, one has to estimate the activity first to conduct the later decoding procedures. According to, the activity level $\hat{K}_a$ can be detected by the power information of the received signal by
\begin{equation}
	\hat{K}_{a} =  \left \lfloor \frac{1}{P}\left( \frac{ \left \| \boldsymbol{Y} \right \|_{F}^2 }{M}-\sigma^2T \right) \right \rceil.
\end{equation}

\subsubsection{Low-Rank Matrix Factorization Via Alternate Minimization}
Considering the low-rank feature of the signal component in \eqref{eq:1}, one can decompose $\boldsymbol{Y}$ into the product of two matrices $\boldsymbol{U}\in \mathbb{C}^{M \times \hat{K}_a}$ and $\boldsymbol{V}\in \mathbb{C}^{\hat{K}_a \times T}$ with random initialization, i.e., $\boldsymbol{Y} \approx \boldsymbol{U}\boldsymbol{V}\approx \boldsymbol{H}\boldsymbol{X}$. Though matrices $\boldsymbol{U}$ and $\boldsymbol{V}$ have identical size of channel matrix $\boldsymbol{H}$ and frame matrix $\boldsymbol{X}$ respectively, they often have different elements and are hoped to produce similar factorization precision towards received signal by decomposition. This decomposition can be formulated as the regularized least square problem below:
\begin{equation}
	\label{eq:4}
	\mathop{\arg\min}\limits_{\boldsymbol{U},\boldsymbol{V}}\|\boldsymbol{Y}-\boldsymbol{U}\boldsymbol{V}\|_{\mathrm{F}}^2+\alpha\|\boldsymbol{U}\|_{\mathrm{F}}^2+\beta\|\boldsymbol{V}\|_{\mathrm{F}}^2,
\end{equation}
where Frobenius norm $\|\boldsymbol{U}\|_{\mathrm{F}}^2$ and $\|\boldsymbol{V}\|_{\mathrm{F}}^2$ are the regularized components preventing overfitting and $0<\alpha,\beta \ll 1$ are positive constants. Let $f_{\boldsymbol{U},\boldsymbol{V}}$ denote the target function in \eqref{eq:4}. One can obtain the gradient of $\boldsymbol{U}$ with fixed $\boldsymbol{V}$ by
\begin{equation}
	\label{eq:5}
	\begin{aligned}
		\frac{\partial f_{\boldsymbol{U},\boldsymbol{V}}}{\partial\boldsymbol{U}}& =\frac{\partial \mathrm{Tr}(\boldsymbol{V}^{\mathrm{H}}\boldsymbol{U}^{\mathrm{H}}\boldsymbol{U}\boldsymbol{V}-2\boldsymbol{Y}^{\mathrm{H}}\boldsymbol{U}\boldsymbol{V})+\alpha \mathrm{Tr}(\boldsymbol{U}^{\mathrm{H}}\boldsymbol{U})}{\partial\boldsymbol{U}}  \\
		&=\frac{\partial \mathrm{Tr}(\boldsymbol{U}^{\mathrm{H}}\boldsymbol{U}\boldsymbol{V}\boldsymbol{V}^{\mathrm{H}}-2\boldsymbol{U}\boldsymbol{V}\boldsymbol{Y}^{\mathrm{H}})+\alpha \mathrm{Tr}(\boldsymbol{U}^{\mathrm{H}}\boldsymbol{U})}{\partial\boldsymbol{U}} \\
		&=2(\boldsymbol{U}\boldsymbol{V}\boldsymbol{V}^{\mathrm{H}}-\boldsymbol{Y}\boldsymbol{V}^{\mathrm{H}}+\alpha\boldsymbol{U}).
	\end{aligned}
\end{equation}
Therefore, the update on $\boldsymbol{U}$ is given by 
\begin{equation}
	\label{eq:6}
	\begin{aligned}
		\frac{\partial f_{\boldsymbol{U},\boldsymbol{V}}}{\partial\boldsymbol{U}} &=2(\boldsymbol{U}\boldsymbol{V}\boldsymbol{V}^{\mathrm{H}}-\boldsymbol{Y}\boldsymbol{V}^{\mathrm{H}}+\alpha\boldsymbol{U})=0 \\
		&\longrightarrow \boldsymbol{U}=\boldsymbol{Y}\boldsymbol{V}^{\mathrm{H}}\left(\boldsymbol{V}\boldsymbol{V}^{\mathrm{H}}+\alpha \boldsymbol{I}\right)^{-1}.
	\end{aligned}
\end{equation}
Similarly, to update $\boldsymbol{V}$, one can obtain the corresponding gradient with fixed $\boldsymbol{U}$ by
\begin{equation}
	\label{eq:7}
	\begin{aligned}
		\frac{\partial f_{\boldsymbol{U},\boldsymbol{V}}}{\partial\boldsymbol{V}} &=\frac{\partial \mathrm{Tr}(\boldsymbol{V}^{\mathrm{H}}\boldsymbol{U}^{\mathrm{H}}\boldsymbol{U}\boldsymbol{V}-2\boldsymbol{Y}^{\mathrm{H}}\boldsymbol{U}\boldsymbol{V})+\beta \mathrm{Tr}(\boldsymbol{V}\boldsymbol{V}^{\mathrm{H}})}{\partial\boldsymbol{V}^{\mathrm{H}}} \\
		&=\frac{\partial \mathrm{Tr}(\boldsymbol{V}\boldsymbol{V}^{\mathrm{H}}\boldsymbol{U}^\mathrm{H}\boldsymbol{U}-2\boldsymbol{V}\boldsymbol{Y}^\mathrm{H}\boldsymbol{U})+\beta \mathrm{Tr}(\boldsymbol{V}\boldsymbol{V}^{\mathrm{H}})}{\partial\boldsymbol{V}^{\mathrm{H}}} \\
		&=2(\boldsymbol{V}^{\mathrm{H}}\boldsymbol{U}^\mathrm{H}\boldsymbol{U}-\boldsymbol{Y}^\mathrm{H}\boldsymbol{U}+\beta\boldsymbol{V}^{\mathrm{H}}),
	\end{aligned}
\end{equation}
and the update on $\boldsymbol{V}$ is given by:
\begin{equation}
	\label{eq:8}
	\begin{aligned}
		\frac{\partial f_{\boldsymbol{U},\boldsymbol{V}}}{\partial\boldsymbol{V}} &=2(\boldsymbol{V}^{\mathrm{H}}\boldsymbol{U}^\mathrm{H}\boldsymbol{U}-\boldsymbol{Y}^\mathrm{H}\boldsymbol{U}+\beta\boldsymbol{V}^{\mathrm{H}})=0 \\
		&\longrightarrow \boldsymbol{V}=\left(\boldsymbol{U}^{\mathrm{H}}\boldsymbol{U}+\beta\boldsymbol{I}\right)^{-1}\boldsymbol{U}^{\mathrm{H}}\boldsymbol{Y}.
	\end{aligned}
\end{equation}
Alternately, one can iteratively update $\boldsymbol{U}$ and $\boldsymbol{V}$ by \eqref{eq:6} and \eqref{eq:8} with random initialization. The complexity is determined by the matrix inversion and product in \eqref{eq:6} and \eqref{eq:8} around $\mathcal{O}(\hat{K}_a^3T+\hat{K}_aTM^2+\hat{K}_aT^2M)$ in order of $\mathcal{O}(\hat{K}_a^3T)$.
%\begin{algorithm}[h]
%	\SetAlgoLined
%	\label{Low-Rank Matrix Completion}
%	\KwIn{Received Signal $\boldsymbol{Y}$}
%	\KwOut{Matrces $\boldsymbol{U}$ and $\boldsymbol{V}$}
%	\textbf{Initialization:}\\
%	$\boldsymbol{U}^0\leftarrow$ Randomly Generated Matrix,$\boldsymbol{V}^0\leftarrow \boldsymbol{0}$\;
%	Convergence Condition: $\tau$\;
%	\While{$\max\left(\|\boldsymbol{U}^q-\boldsymbol{U}^{q-1}\|_2^2,\|\boldsymbol{V}^q-\boldsymbol{V}^{q-1}\|_2^2\right)\ge\tau$}{
%		$\boldsymbol{V}\leftarrow \left(\boldsymbol{U}^{\mathrm{H}}\boldsymbol{U}+\beta\boldsymbol{I}\right)^{-1}\boldsymbol{U}^{\mathrm{H}}\boldsymbol{Y}$ \eqref{eq:8}\;
%		$\boldsymbol{U}\leftarrow \boldsymbol{Y}\boldsymbol{V}^{\mathrm{H}}\left(\boldsymbol{V}\boldsymbol{V}^{\mathrm{H}}+\alpha \boldsymbol{I}\right)^{-1}$ \eqref{eq:6};
%	}
%	\caption{Low-Rank Matrix Completion \\ Via Alternate Regularized Least Square Minimization }
%\end{algorithm}
\subsubsection{Ambiguity Compensation Via Codeword Sparsity}
It should be noted that the low-rank matrix factorization by \eqref{eq:6} and \eqref{eq:8} are not unique but randomized by the initialization. After decomposition, one can obtain ambiguous channel matrx $\boldsymbol{U}=\boldsymbol{H}\boldsymbol{G}$ and ambiguous frame matrix $\boldsymbol{V}=\boldsymbol{G}^{-1}\boldsymbol{X}$, i.e., $\boldsymbol{UV}=\boldsymbol{HX}$. The influence from ambiguity matrix $\boldsymbol{G} \in \mathbb{C}^{\hat{K}_a \times \hat{K}_a}$ needs to be compensated before one can obtain the true channel matrix and frame matrix. To tackle such issue, we exploit the sparsity of active on-off patterns. The decomposed matrix $\boldsymbol{V}$ during the time of the first $T_p$ slots can be written by
\begin{equation}
	\label{eq:9}
	\begin{aligned}
		\boldsymbol{V}(:,1:T_p)\in \mathbb{C}^{\hat{K}_a \times T_p}&=\boldsymbol{G}^{-1}\boldsymbol{X}(:,1:T_p) \\
		&=\boldsymbol{G}^{-1}\left[\boldsymbol{a}_{i_1}^{\mathrm{T}},\boldsymbol{a}_{i_2}^{\mathrm{T}},\ldots,\boldsymbol{a}_{i_{\hat{K}_a}}^{\mathrm{T}}\right]^{\mathrm{T}},
	\end{aligned}
\end{equation}
where $\boldsymbol{X}(:,1:T_p)$ has only $\hat{K}_a$ out of $2^{B_1}$ codewords in $\boldsymbol{A}$ and $\hat{K}_a \ll 2^{B_1}$. Therefore, the estimation on the ambiguity matrix can be converted into a CS problem by:
\begin{equation}
	\label{eq:10}
	\boldsymbol{V}(:,1:T_p)^{\mathrm{T}}\in \mathbb{C}^{T_p \times \hat{K}_a} \approx \boldsymbol{A}\tilde{\boldsymbol{G}},
\end{equation}
where $\tilde{\boldsymbol{G}} \in \mathbb{C}^{2^{B_1}\times \hat{K}_a}$ is a row sparse matrix with only $\hat{K}_a$ nonzero rows. These $\hat{K}_a$ nonzero rows constitute the ambiguity matrix to be estimated. We solve this CS problem by the simultaneous orthogonal matching pursuit (SOMP) algorithm.
%Algorithm \ref{alg:SOMP_MIMO}.
%\begin{algorithm}[h]
%	\SetAlgoLined
%	\label{alg:SOMP_MIMO}
%	\KwIn{Matrix $\boldsymbol{V}(:,1:T_p)$, codebook $\boldsymbol{A}$}
%	\KwOut{Matrix $\tilde{\boldsymbol{G}}$}
%	Initialization: $k\leftarrow 0$, $\boldsymbol{R}\leftarrow \boldsymbol{V}(:,1:T_p)$, $\mathcal{S}\leftarrow \emptyset$ \;
%	\While{$k \le \hat{K}_a$}{
%		\textcolor{blue}{$i_{k}\leftarrow \mathop{\arg\max}\limits_{i_k\in\{1,2,\ldots,2^{B_1}\}}\frac{\|\boldsymbol{R}^{\mathrm{H}}\boldsymbol{A}(:,i_k)\|_2}{\boldsymbol{A}(:,i_k)}$}\;
%		$\mathcal{S}\leftarrow \mathcal{S} \cup i_{k}$\;
%		$\boldsymbol{\Xi }\leftarrow \boldsymbol{A}\left(:,\{\mathcal{S}\}\right)$ \;
%		$\boldsymbol{R}\leftarrow \left(\boldsymbol{R}-\boldsymbol{\Xi }\boldsymbol{\Xi }^{\dagger}\boldsymbol{V}(:,1:T_p)\right )$ and $\left(\cdot\right)^{\dagger}$ denotes Moore-Penrose inverse\;
%		$k\leftarrow k+1$\;}
%	$\tilde{\boldsymbol{G}}\leftarrow \boldsymbol{A}\left(:,\{\mathcal{S}\}\right)^{\dagger}\boldsymbol{V}(:,1:T_p).$
%	\caption{Ambiguity Compensation Via Codeword Sparsity with SOMP}
%\end{algorithm}

SOMP finds the most correlated codeword in $\boldsymbol{A}$ with the residual matrix $\boldsymbol{R}$ initialized as $\boldsymbol{V}(:,1:T_p)$, by $\mathop{\arg\max}\limits_{i_k\in\{1,2,\ldots,2^{B_1}\}}\frac{\|\boldsymbol{R}^{\mathrm{H}}\boldsymbol{A}(:,i_k)\|_2}{\boldsymbol{A}(:,i_k)}$. The support set $\mathcal{S}$ restores the indices of the detected codewords. Then, the projection matrix $\boldsymbol{\Xi}$ is constructed by the detected codewords $\boldsymbol{A}\left(:,\{\mathcal{S}\}\right)$ and the residual is updated by $\left(\boldsymbol{R}-\boldsymbol{\Xi }\boldsymbol{\Xi }^{\dagger}\boldsymbol{V}(:,1:T_p)\right )$. Finally, the ambiguity matrix is estimated by $\boldsymbol{A}\left(:,\{\mathcal{S}\}\right)^{\dagger}\boldsymbol{V}(:,1:T_p)$ and $\left(\cdot\right)^{\dagger}$ denotes Moore-Penrose inverse. The estimated ambiguity matrix is given by $\boldsymbol{G} = \left(\tilde{\boldsymbol{G}}(\{\mathcal{S}\},:)^{\mathrm{T}}\right)^{-1}$.
%\begin{equation}
%	\label{eq:11}
%	\boldsymbol{G} = \left(\tilde{\boldsymbol{G}}(\{\mathcal{S}\},:)^{\mathrm{T}}\right)^{-1}
%\end{equation}
With knowledge of $\boldsymbol{G}$, the channel matrix and the frame matrix can be obtained by $	\boldsymbol{H} = \boldsymbol{U}\boldsymbol{G}^{-1}, \boldsymbol{X} = \boldsymbol{G}\boldsymbol{V}$.
%\begin{equation}
%	\label{eq:12}
%	\boldsymbol{H} = \boldsymbol{U}\boldsymbol{G}^{-1}, \boldsymbol{X} = \boldsymbol{G}\boldsymbol{V}
%\end{equation}
The complexity of SOMP can be reduced into $M2^{B_1}\log2^{B_1}$ if sub-sampled Discrete Fourier Transform (DFT) matrix is utilized with matrix multiplication alternated into fast Fourier transformation (FFT)\cite{Slotted_Pilot}.
\subsubsection{Low Complexity Joint Pattern and Data Detection}
After obtaining the channel estimation, the proposed decoder conducts a joint pattern and data detection (JPDD) in the spirit of approximate message passing (AMP). The signals to be detected $\boldsymbol{Y}_c$ is during the time of $[T_p+1:T]$, i.e, $\boldsymbol{Y}_c=\boldsymbol{Y}\left(:,T_p+1:T\right)$. The slot-wise signal model is given by:
\begin{equation}
	\label{eq:13}
	\boldsymbol{y}_t=\boldsymbol{H}\boldsymbol{x}_t+\boldsymbol{n}_t,t\in\{T_p+1,T_p+2,\ldots,T\},
\end{equation}
where $\boldsymbol{n}_t$ is the AWGN in \eqref{eq:1}. The states of elements in $\boldsymbol{x}_t$ are to be detected with the observations and the estimated channel. For ODMA, the signal at single slot has states of $\{0,\mathcal{Q}\}$ where 0 means idle state. The goal is to calculate the a posteriori probabilities (APPs) of elements in $\boldsymbol{x}_t$ towards all states. Let $p(x_{t,j}=s|\boldsymbol{y}_t,s\in \{0,\mathcal{Q}\})$ denote the APP of $j$-th element in $\boldsymbol{x}_t$ based on the noisy observations. 

Different to the detection algorithm in \cite{IOTJ_ODMA} and inspired by \cite{AMP1,AMP2}, a JPDD algorithm with lower complexity is proposed. The iteration-wise calculation steps are listed as follows:
\begin{equation*}
	\begin{aligned}
		\boldsymbol{s}^{q}&
		=\tilde{\boldsymbol{x}}_t^q+\frac{1}{M}\boldsymbol{H}^{\mathrm{H}}\boldsymbol{r}^q\\
		\tilde{\boldsymbol{x}}_t^{q+1}&
		=\eta\left[\boldsymbol{s}^{q},\sigma^2(\frac{1}{M}+\tau^q)\right]\\
		\tau^{q+1}&
		=\frac{\beta}{\sigma^2}\left \langle \zeta \left[\boldsymbol{s}^q,\sigma^2(\frac{1}{M}+\tau^q)\right] \right \rangle \\
		\boldsymbol{r}^{q+1}&=\boldsymbol{y}_t-\boldsymbol{H}\tilde{\boldsymbol{x}}_t^{q+1}+\frac{\tau^{q+1}}{1+\tau^q}\boldsymbol{r}^q
	\end{aligned}
\end{equation*}
where $q$ refers to the index of the $q$-th iteration, $\beta=\frac{\hat{K}_a}{M}$, the initialization setups are  $\boldsymbol{x}_t^{1}=\boldsymbol{0}$ or equalized version, $\boldsymbol{r}^1=\boldsymbol{y}_t, \tau^1=\frac{\mathrm{Var}[s]}{\sigma^2}$, and $\left \langle \cdot \right \rangle $ denotes $\left \langle \boldsymbol{x} \right \rangle=\frac{1}{N}\sum_{i=1}^{N}x_i$. Meanwhile, functions $\eta[x,z]$ and $\zeta[x,z]$ refer to the element-wise expectation and variance of via APPs toward different constellation symbols, i.e., $\eta[x,z]=\int_{s\in \{0,\mathcal{Q}\}}s\cdot p(x=s|s,z)\mathrm{d}x, \zeta[x,z]=\int_{s\in \{0,\mathcal{Q}\}}s^2\cdot p(x=s|s,z)\mathrm{d}x-|\eta[x,z]|^2$ and the iteration-wise APP update can be done through different euclidean distance between the $\boldsymbol{s}^{q}$, i.e., $\frac{\exp{\left(\frac{-|\boldsymbol{s}^{q}-s_i|^2}{2\tau^q}\right)}}{\sum\exp{\left(\frac{-|\boldsymbol{s}^{q}-s_i|^2}{2\tau^q}\right)}},s_i\in \{0,\mathcal{Q}\}$. Each iteration consumes complexity around $\mathcal{O}(M\hat{K}_a)$ and thus overall detection takes up $\mathcal{O}(M\hat{K}_aT_c)$ complexity per round, i.e., linear to the number of active users which is much smaller than the detector in \cite{IOTJ_ODMA} whose complexity scales as $\mathcal{O}(M\hat{K}^2_aT_c)$.

After verdict by APPs, the on-off pattern is retrieved by selecting the pattern codewords with maximum probability multiplication from each non-zero elements. Simultaneously, the log-likelihood ratio (LLR) of FEC encoded bits can be calculated to conduct FEC decoding. For example, 
%if BPSK is adopted, the LLR is calculated by
%\begin{equation}
%	\label{eq:37}
%	llr_i=\log\frac{p_i(s=s_1|\tilde{\boldsymbol{y}}_t,\{0,\mathcal{Q}\})}{p_i(s=s_2|\tilde{\boldsymbol{y}}_t,\{0,\mathcal{Q}\})},
%\end{equation}
for quadrature phase shift keying (QPSK), since two bits are projected into one constellation symbol, i.e., $[b_0,b_1] \rightarrow s_{b_0,b_1}$, the LLRs of $b_0$ and $b_1$ can be obtained respectively by:
\begin{equation}
	\label{eq:38}
	\begin{aligned}
		llr_0&=\log\frac{p_i(s_{0,0}|\tilde{\boldsymbol{y}}_t,\{0,\mathcal{Q}\})+p_i(s_{0,1}|\tilde{\boldsymbol{y}}_t,\{0,\mathcal{Q}\})}{p_i(s_{1,0}|\tilde{\boldsymbol{y}}_t,\{0,\mathcal{Q}\})+p_i(s_{1,1}|\tilde{\boldsymbol{y}}_t,\{0,\mathcal{Q}\})}, \\ llr_1&=\log\frac{p_i(s_{0,0}|\tilde{\boldsymbol{y}}_t,\{0,\mathcal{Q}\})+p_i(s_{1,0}|\tilde{\boldsymbol{y}}_t,\{0,\mathcal{Q}\})}{p_i(s_{0,1}|\tilde{\boldsymbol{y}}_t,\{0,\mathcal{Q}\})+p_i(s_{1,1}|\tilde{\boldsymbol{y}}_t,\{0,\mathcal{Q}\})}.
	\end{aligned}
\end{equation}
\subsubsection{Successive Interference Cancellation (SIC)}
In this part, SIC is adopted as an iterative approach to generate system performance gain. After FEC decoding, the decoder will produce a preliminary message list. The decoder first conducts parity check on messages in $\tilde{\mathcal{L}}$ and then subtracts the re-encoded-and-modulated signals from the original signals in \eqref{eq:1} with refined channel estimation by 
	\begin{equation}
		\boldsymbol{H}_{\mathrm{passed}}=\left(\boldsymbol{X}_{\mathrm{passed}}^{\mathrm{H}}\boldsymbol{X}_{\mathrm{passed}}+\sigma^2\boldsymbol{I}_{|\mathcal{L}_{\mathrm{passed}}|}\right)^{-1}\boldsymbol{X}_{\mathrm{passed}}^{\mathrm{H}}\boldsymbol{Y},
	\end{equation}	
	where $\boldsymbol{X}_{\mathrm{passed}}$ is the re-encoded and re-modulated vectors, and $\mathcal{L}_{\mathrm{passed}}$ denote the set of parity check passed messages. The updates for the next decoding round are $\boldsymbol{Y}\leftarrow \boldsymbol{Y}-\boldsymbol{X}_{\mathrm{passed}}\boldsymbol{H}_{\mathrm{passed}}$ and $\tilde{\mathcal{L}} \leftarrow \tilde{\mathcal{L}} \cup \mathcal{L}_{\mathrm{passed}}$. The SIC procedure continues until there are no messages who can pass the parity check or the decoding reaches a prescribed SIC upper-bound.
	\begin{figure}[htp]
		\centering
		\includegraphics[width=3in]{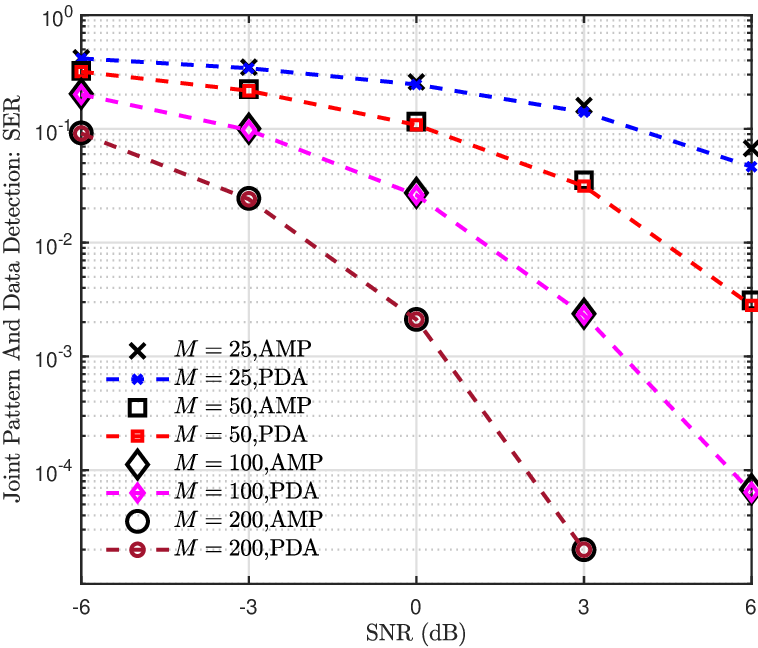}
		\caption{Symbol error rate (SER) comparison between the proposed JPDD-AMP detector and the JPDD-PDA detector in \cite{IOTJ_ODMA} under different $\mathrm{SNR}$ (dB) and number of antenna $M$.}
		\label{SER}
	\end{figure}
	\begin{figure}[htp]
		\centering
		\includegraphics[width=3in]{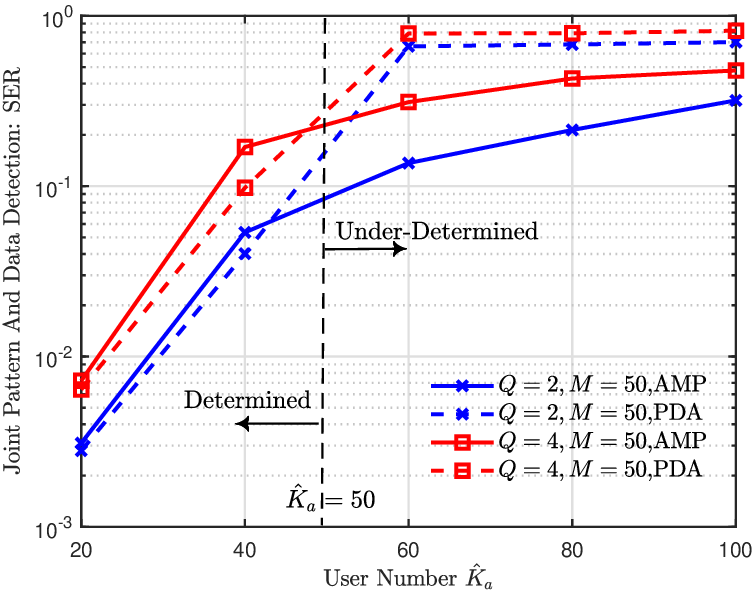}
		\caption{SER comparison in cases of determined and under-determined estimation with $\mathrm{SNR}=5$dB, $M=50$ and different modulation order $Q=|\mathcal{Q}|$ for $\{0,\mathcal{Q}\}$.}
		\label{SER2}
	\end{figure}
\section{Numerical Results}\label{sec.5}
\begin{figure*}[htp] 
	\centering
	\subfloat[]{\includegraphics[width=2.35in]{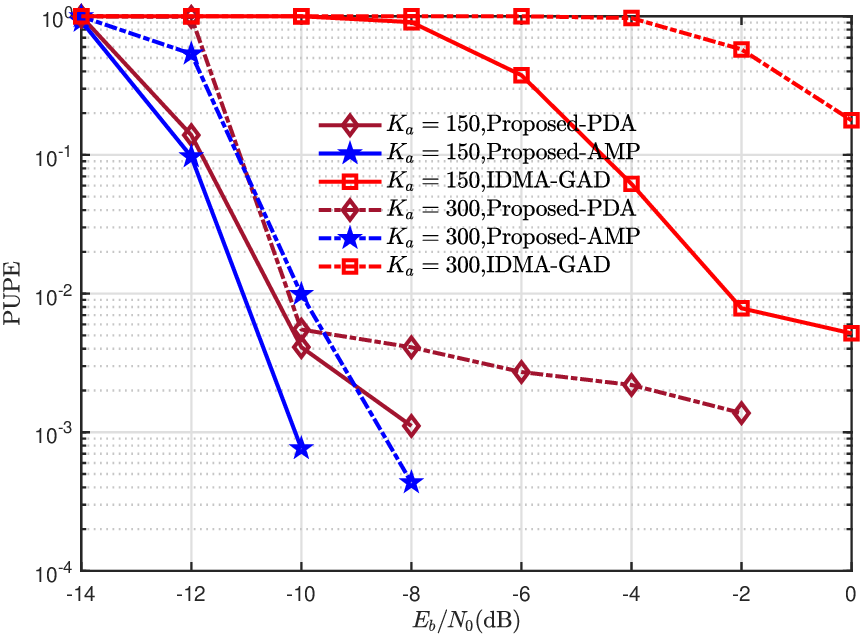}%
		\label{VS_IDMA_BNR}}
	\hspace{1mm}
	\subfloat[]{\includegraphics[width=2.35in]{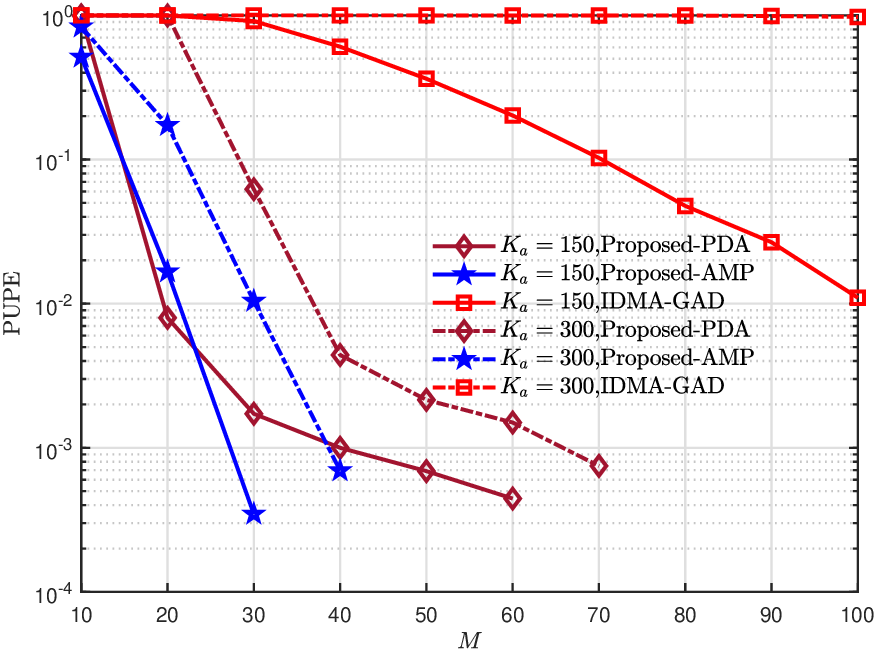}%
		\label{VS_IDMA_M}}
	\hfil
	\subfloat[]{\includegraphics[width=2.3in]{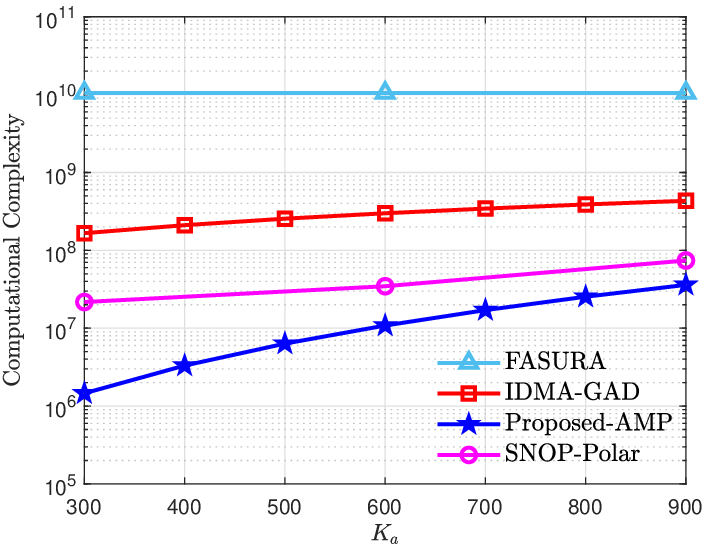}%
		\label{complexity}}
	\caption{PUPE Comparisons between the proposed scheme with IDMA and computational complexity comparison with recent works: a) PUPE versus energy-per-bit $E_b/N_0$(dB) with $M=50$; b) PUPE versus the number of antenna $M$ under $E_b/N_0=-6$dB; c) Complexity versus number of active users $K_a$.}
	\label{PUPE_M_BNR}
\end{figure*}
In this section, we illustrate the performance of the proposed scheme compared with state-of-the-arts including pilot-coupled ODMA\cite{ODMA5}, pilot-free ODMA\cite{TWC_ODMA}, IDMA\cite{SPARC-LDPC}, FASURA\cite{FASURA}, SKP\cite{SKP}, Pilot-based URA\cite{Collision_Probability}, slotted no-orthogonal pilot-based URA with Polar code (SNOP-Polar)\cite{Slotted_Pilot} and multi-stage orthogonal URA with Polar code (MSOP-Polar)\cite{Multi_Stage_1} and the quasi-static MIMO URA bound in \cite{Bound_with_Quasi_Channel}. Here are some universal parameter setups: The total channel uses is $L=3200$, the number of antenna is $M=50$, the number of transmission chunk equals to $J=16$, i.e., $T=L/J=200$. The bits allocation is $B_{\mathrm{chunk}}=4$, $B_1=14$, $B_2=14$, $B_3=68$. Polar code with $128$ code length with 14-CRC is adopted. The CRC generation polynomial is $x^{14}+x^{10}+x^9+x^7+x^6+x^5+x+1$. The size of list decoding equals to 128. The encoded bits are modulated with QPSK. For $K_a\le 600$, $T_p=50,T_c=150$ and for $K_a\ge 800$, $T_p=75$, $T_c=125$. After trails, the power allocation ratio is recommended as $\alpha=0.2$. For IDMA, it takes up identical channel uses including a CS phase. For $K_a=150$, the length of CS phase is fixed to $200$ and for $K_a=300$, the length of CS phase is $400$.
\subsubsection{Detection Performance Comparison}
In this section, the performance of the proposed AMP-oriented JPDD detector and the existing probabilistic data association (PDA)-based detector\cite{IOTJ_ODMA} is compared. Model \eqref{eq:13} is adopted with assumption of known channel coefficients and transmitted signals randomly generalized from $\{0,\mathcal{Q}\}$. The signal-to-noise ratio (SNR) is defined by: $\mathrm{SNR} = \frac{\mathbb{E}\left\{\|\boldsymbol{H}\boldsymbol{x}_t\|_2^2\right\}}{\mathbb{E}\left\{\|\boldsymbol{n}_t\|_2^2\right\}}$.
%\begin{equation}
%	\mathrm{SNR} = \frac{\mathbb{E}\left\{\|\boldsymbol{H}\boldsymbol{x}_t\|_2^2\right\}}{\mathbb{E}\left\{\|\boldsymbol{n}_t\|_2^2\right\}}.
%\end{equation}

In Fig. \ref{SER}, the symbol error rate (SER) of slot-wise JPDD model \eqref{eq:13} is illustrated with $\hat{K}_a=25$ under different number of receiving antennas $M\in \{25,50,100,200\}$ and SNR. The detected signals are verdict by maximum APP. Overall, the proposed JPDD-AMP detector has marginal performance loss compared wtih PDA detector under large number of receiving antenna but with much reduced complexity, which is a favorable feature.

Particularly, both estimation cases of determined ($\hat{K}_a\le M$) and under-determined ($\hat{K}_a > M$) are investigated in Fig. \ref{SER2} with fixed $\mathrm{SNR}=5$dB and $M=50$. Before $\hat{K}_a=50$, i.e., determined case, the observation dimension adequately support the JPDD task and the performance difference is marginal between the proposed AMP detector and the PDA detector. However, after $\hat{K}_a=50$, i.e., under-determined case, the performance of PDA detector deteriorates drastically whereas the proposed AMP JPDD detector illustrates much slower inertial and produces nearly $50\%$ better performance, which is an appealing feature in terms of massive connectivity.
\begin{figure}[htp]
	\centering
	\includegraphics[width=3in]{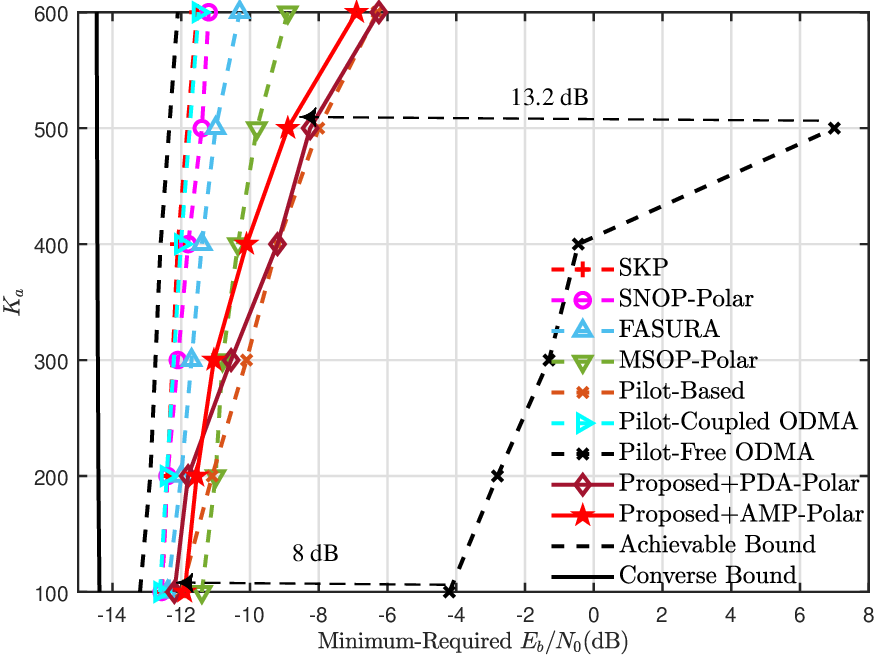}
	\caption{Minimum required $E_b/N_0$ (dB) to reach PUPE $P_e=0.05$ under $L=3200$, $M=50$: The capacity performance of the pilot-free receiver is significantly enhanced, ranging from 8 dB to 13.2 dB.}
	\label{system_capacity}
\end{figure}

\subsubsection{PUPE Performance with Slotted Transmission}
In Fig. \ref{PUPE_M_BNR}, the PUPE performance under various setups with IDMA via Gaussian approximation detector (GAD) with benchmark.
	
Fig. \ref{VS_IDMA_BNR} illustrates the PUPE performance of the proposed scheme and IDMA versus different energy-per-bit $E_b/N_0$(dB) under different number of active devices $K_a=150$ and $K_a=300$ with $M=50$. Both schemes shows a water-falling performance with increased energy-per-bit. However, the proposed scheme needs much less $E_b/N_0$ to enable the decoding procedure. Meanwhile, the proposed with AMP-based detector shows faster PUPE drop compared with PDA detector\cite{IOTJ_ODMA}. Again, this demonstrates the proposed AMP JPDD detector shows more tolerance for the under-determined estimation case.
	
Fig. \ref{VS_IDMA_M} illustrates the PUPE versus different number of antennas with $E_b/N_0=-6$dB. The proposed slotted ODMA with both PDA and AMP detector produce superior performance compared with IDMA. Moreover, the proposed with AMP detector also illustrates water-falling PUPE drop and a better performance at low antenna amount region compared with proposed with PDA detector which derives from the enhanced under-determined estimation ability illustrated in Fig. \ref{SER2}.
% \usepackage{multirow}
% \usepackage{multirow}
\begin{table*}[htp]
	\centering
	\caption{Computational Complexity and Capacity Trade-Offs}
	\label{tab:trade-off}
	  	\scalebox{0.8}{
	  		\renewcommand{\arraystretch}{1.15}
	\begin{tabular}{cc|cc|cc}
		\hline
		\multicolumn{2}{c|}{$M=50$}                                                                                                         & \multicolumn{2}{c|}{Complexity $\times 10^{6}$}  & \multicolumn{2}{c}{Minimum-Required $E_b/N_0$ (dB)} \\ \hline
		\multicolumn{2}{c|}{Scheme}                                                                                                         & $K_a=300$                  & $K_a=600$           & $K_a=300$                    & $K_a=600$            \\ \hline
		\multicolumn{2}{c|}{FASURA}                                                                                                         & 10486                      & 10486               & -11.7                        & -10.3                \\
		\multicolumn{2}{c|}{SNOP-Polar}                                                                                                     & 21.86                      & 34.6                & -12.1                        & -11.2                \\
		\multicolumn{2}{c|}{Proposed-AMP}                                                                                                   & 1.4592                     & 10.828              & -11.05                       & -6.9                 \\ \hline
		\multicolumn{2}{c|}{Gap Between Schemes}                                                                                            & \multicolumn{2}{c|}{Complexity Gap (dB)}         & \multicolumn{2}{c}{Capacity Gap (dB)}               \\ \hline
		\multicolumn{2}{c|}{Scheme}                                                                                                         & $K_a=300$                  & $K_a=600$           & $K_a=300$                    & $K_a=600$            \\ \hline
		\multicolumn{2}{c|}{FASURA}                                                                                                         & 38.565 dB                  & 29.8606 dB          & 0.35 dB                      & 3.4 dB               \\
		\multicolumn{2}{c|}{SNOP-Polar}                                                                                                     & 11.7554 dB                 & 5.0453 dB           & 1.05 dB                      & 4.3 dB               \\ \hline
		\multicolumn{1}{c|}{\multirow{2}{*}{\begin{tabular}[c]{@{}c@{}}Trade-Off:\\ Complexity Gap-Capacity Gap\end{tabular}}} & FASURA     & \multirow{2}{*}{$K_a=300$} & \textbf{38.215 dB}  & \multirow{2}{*}{$K_a=600$}   & \textbf{26.4606 dB}  \\ \cline{2-2}
		\multicolumn{1}{c|}{}                                                                                                  & SNOP-Polar &                            & \textbf{10.7054 dB} &                              & \textbf{0.7453 dB}   \\ \hline
		\end{tabular}}
\end{table*}
\subsubsection{Capacity Performance and Complexity Trade-Off}
Fig. \ref{system_capacity} illustrates the minimum required energy-per-bit to reach 0.05 PUPE of various state-of-the-arts with the achievable bound and converse bound under channel uses of $L=3200$ and number of antennas $M=50$. With up to 300 active users, the proposed scheme exhibits performance close to SKP and SNOP-Polar, with a minimum energy-per-bit gap of 0.37 dB and 0.2 dB, respectively, compared to FASURA. Additionally, the proposed scheme outperforms both the MSOP-Polar and Pilot-Based schemes. Compared to the existing pilot-free ODMA scheme, the proposed scheme provides a capacity gain ranging from 8 dB to 13.2 dB. It significantly reduces the performance gap between pilot-free ODMA and pilot-coupled ODMA.

Moreover, the proposed scheme with the AMP detector achieves enhanced capacity performance in the high-activity region compared to the PDA detector. This improvement stems from the better detection ability in the under-determined case. Complexities of different schemes are illustrated in Fig. \ref{complexity} and the trade-offs between computational complexity and systematic capacity are summarized in Table \ref{tab:trade-off}. A positive trade-off can be observed between the proposed scheme and recent works.
%
%\begin{figure}[htp]
%	\centering
%	\includegraphics[width=3.5in]{complexity.eps}
%	\caption{Complexity comparison with state-of-the-arts.}
%	\label{complexity}
%\end{figure}
\section{Conclusions}\label{sec.6}
This work explores receiver designs applicable to both pilot-uncoupled and pilot-free ODMA for URA in MIMO systems, addressing the challenges of complexity reduction and capacity enhancement. For the JPDD problem, the proposed AMP detector reduces processing complexity from quadratic to linear order compared to existing methods. Furthermore, the proposed AMP-JPDD approach demonstrates more robust performance in under-determined linear regression models. An enhancement of up to 13 dB in capacity performance is observed compared to existing pilot-free ODMA, using a similar receiver design based on matrix factorization. Future research will focus on asynchronous transmission design and adaptations for more advanced MIMO systems.
\balance

%\newpage
% your own custom macro containing the $\backslash${\tt{includegraphics}} command to make things
%\bf{If you include a photo:}\vspace{-33pt}
%\begin{IEEEbiography}[{\includegraphics[width=1in,height=1.25in,clip,keepaspectratio]{fig1}}]{Michael Shell}
%Use $\backslash${\tt{begin\{IEEEbiography\}}} and then for the 1st argument use $\backslash${\tt{includegraphics}} to declare and link the author photo.
%Use the author name as the 3rd argument followed by the biography text.
%\end{IEEEbiography}

\vfill

\end{document}